\documentclass[a4paper]{article}

\usepackage{INTERSPEECH2016}

\usepackage{graphicx}
\usepackage{amssymb,amsmath,bm}
\usepackage{textcomp}
\usepackage{epstopdf}
\usepackage{tabularx}
\usepackage{multirow}
\usepackage{subcaption}
\usepackage{floatrow}
\usepackage{url}
\def\vec#1{\ensuremath{\bm{{#1}}}}

\newfloatcommand{capbtabbox}{table}[][\FBwidth]

\ninept

\title{Features and Kernels for Audio Event Recognition}

\makeatletter
\def\name#1{\gdef\@name{#1\\}}
\makeatother \name{{\em Anurag Kumar, Bhiksha Raj}}

\address{Carnegie Mellon University, Pittsburgh, PA, USA - 15217 \\
  {\small \tt alnu@andrew.cmu.edu, bhiksha@cs.cmu.edu}
}

\begin{document}

  \maketitle
  %
\begin{abstract}
One of the most important problems in audio event detection research is absence of benchmark results for comparison with any proposed method. Different works consider different sets of events and datasets which makes it difficult to comprehensively analyze any novel method with an existing one. In this paper we propose to establish results for audio event recognition on two recent publicly-available datasets. In particular we use Gaussian Mixture model based feature representation and combine them with linear as well as non-linear kernel Support Vector Machines.

\end{abstract}
\vspace{-0.15in}
\section{Introduction}
\label{sec:intro}
\vspace{-0.1in}
In recent years automatic content analysis of audio recordings has been gaining attention among the audio research community. The goal is to develop methods which can automatically detect the presence of different kinds of audio events in a recording. Audio event detection (AED) research is driven by its application in several areas. These include areas such as multimedia information retrieval or multimedia event detection \cite{2015trecvidover} where the audio component contains important information about the content of the multimedia data. This is particularly important for supporting content-based search and retrieval of multimedia data on the web. Other applications such as surveillance \cite{1}, wildlife monitoring \cite{stowell2014} \cite{ruizmultiple}, context aware systems \cite{battaglino2015} \cite{eronen}, health monitoring etc. are also motivating audio event detection research. A variety of methods have been proposed for AED in the last few years. A GMM-HMM structure, similar to that used in automatic speech recognition was presented in \cite{zhuang2010}. A simple yet effective approach is the bag of words representation \cite{7}\cite{supbow} \cite{kumar2013event}. This approach also features in audio event detection in noisy environments \cite{foggia2015reliable}.  Given the complexity of audio event detection, deep neural networks which are known for modeling highly non-linear functions have also been explored for AED \cite{Ashraf2015} \cite{gencoglu}. Some other interesting approaches involve use of acoustic unit descriptors for AED \cite{12}, spectral exemplars for AED \cite{specex}, and matrix factorization of spectral representations \cite{mesaros2015sound}. Lack of supervised data for audio events has also led to some weakly supervised audio event detection works \cite{kumar2016audio} \cite{anuragweakly}. 

All of the methods have been shown to be effective and reasonable upto an extent on datasets and events on which they were evaluated. However, a major problem is that we fail to understand how they compare against each other. Most of them use different datasets or sets of sound events and given the variations we encounter across datasets and more importantly across sound events it is difficult to assess different methods. Another factor lacking in several of the current works is analysis of the proposed method on a reasonably large vocabulary of sound events. Usually, only small sets of audio events are considered. The absence of publicaly available An important reason for these concerns in current AED literature is the absence of publicly available standard sound event databases which can serve as a common testbed for any proposed method. More importantly, the dataset should be of reasonable size in terms of total duration of audio and in terms of the vocabulary of events. Open-source dataset helps in standardization of research efforts on a problem and give a better understanding of results shown by any proposed method. For example, in image recognition and classification results are prominently reported and compared on well known standard datasets such as PASCAL \cite{pascal-voc-2011}, CIFAR \cite{krizhevsky2009learning} and ImageNet\cite{deng2009imagenet}. 

Recently, the audio community have attempted to address this problem by creating standard sound event databases meant for audio event detection tasks \cite{salamon2014dataset} \cite{piczak2015esc} \cite{mesaros2016tut}. The \emph{UrbanSounds8k} \cite{salamon2014dataset} dataset contains $10$ different sound events with a total of about $8.75$ hours of data. The \emph{ESC-50} \cite{piczak2015esc} dataset contains labeled data for $50$ sound events and a total of $2.8$ hours of audio. An important goal of our work is to perform a comprehensive analysis of these two datasets for audio event detection. These two public datasets addresses event detection concerns in terms of both total duration and vocabulary of sound events. The total duration of the audio is reasonably large and overall they allowed us to present results on $56$ unique sound events. Specifically, we analyze Gaussian Mixture Models (GMMs) for feature representation. We inspect two forms of feature representations using GMMs.  The first one is a \emph{soft-count} histogram representation obtained using a GMM and the second uses maximum {\em a posteriori} (MAP) adaptation of GMM parameters to obtain a fixed-dimensional  feature representation for audio segments. The first feature is similar to the bag of audio words representation. The second feature obtained through MAP adaptation tries to capture the actual distribution of low-level features of a recording. On the classifier side we use Support Vector Machines (SVMs) where we explore different kernels for audio event detection. 

The ESC-50 dataset also allowed us to study audio event detection at different hierarchies. The $50$ events of ESC-50 broadly belong to $5$ different higher semantic categories. We report and analyze results for both lower-level events and higher semantic categories.  

It is important to note that our analysis on these two datasets is very different from the simple multi-class classification analysis done by the authors of these datasets. In particular, we are interested in audio event detection where our goal is to detect presence or absence (binary) of an event in a given recording. Our analysis is in terms of ranked measure (AP) for each event as well as area under detection error curves for each event. 

The rest of the paper is organized as follows. In Section 2 we provide a brief description of ESC-50 and UrbanSounds8k datasets. We describe the features and kernels used for audio event detection in Section 3 and report results in Section 4. Section 5 concludes our work. 
\section{Datasets}
\vspace{-0.05in}
\label{sec:dtst}
\textbf{UrbanSounds8k}: This dataset contains a total of $8732$ audio clips over $10$ different sound event classes. The event classes are \emph{air conditioner (AC), car horn (CH), children playing (CP), dog barking (DB), drilling (DR), engine idling (EI), gun shot (GS), jackhammer (JK), siren (SR)} and \emph{street music (SM)}. The total duration of the recordings is about $8.75$ hours. The length of each audio clip is less than or equal to $4$ seconds. The dataset comes with a presorted $10$ folds and we use the same fold structure in our experiments. \\
\textbf{ESC-50}: ESC-50 contains a total of $2000$ audio clips over $50$ sound events. Examples sound classes are \emph{dog barking, rain, crying baby, fireworks, clock ticking} etc. The full list of sound events can be found in \cite{piczak2015esc}. The entire set of $50$ events in the dataset belong $5$ groups, namely , \emph{animals (e.g dog, cat), natural soundscapes (e.g rain, seawaves), non-speech human sounds (e.g sneezing, clapping), interior domestic sounds (e.g door knock, clock), and exterior sounds (e.g helicopter, siren)}. The total duration of the recordings in this dataset is around $2.8$ hours. We divide the dataset into $10$ folds for our experiments. For future use of our setup by others, the details of this fold structure is available on this webpage \cite{aedres}. 

For both datasets we convert all audios into single channel and resample all audio to $44.1$KHz sampling frequency. 
\vspace{-0.1in}
\section{Features and Kernels}
\vspace{-0.1in}
\subsection{Features}
\vspace{-0.05in}
Before we can go into the details of different feature representation we need low-level characterization of audio signals. We use Mel-Frequency Cepstral Coefficients (MFCC) as low-level representations of audio signals. Let these $D$-dimensional MFCC vectors for a recording be represented as $\vec{x}_t$, where $t=1\,\,to\,\,T$, $T$ being the total number of MFCC vectors for the recording. 

Broadly, we employ two higher level feature representations for characterizing audio events. Both of these features attempt to capture the distribution of MFCC vectors of a recording. We will refer to these as $\vec{\alpha}$ and $\vec{\beta}$ features and the sub-types will be represented using appropriate sub-scripts and superscripts. 

The first step to obtain higher-level fixed dimensional representation is to train a GMM on MFCC vectors of the training data. Let us represent this GMM by $\mathcal{G} = \{w_k,N(\vec{\mu}_k, \Sigma_k), k = 1 \,\,to \,\,M\}$, where $w_k$, $\vec{\mu}_k$ and $\Sigma_k$ are the mixture weight, mean and covariance parameters of the $k^{th}$ Gaussian in $\mathcal{G}$. We will assume diagonal covariance matrices for all Gaussians and $\vec{\sigma}_k$ will represent the diagonal vector of $\Sigma_k$. 
\subsubsection{\vec{\alpha} features}
Our first feature is based on the bag of audio words \cite{7} representation where we employ a GMM for a more robust representation \cite{kumar2013event} \cite{foggia2015reliable}. We will refer to these representations as $\vec{\alpha}$ feature representations.  Given the MFCC vectors $\vec{x}_t$ of a recording, we compute the probabilistic assignment of $\vec{x}_t$ to the $k^{th}$ Gaussian as in Eq \ref{eq:postr}. 
\begin{align}
Pr(k | \vec{x}_{t}) = & \frac{w_{k}N(\vec{x}_{t} ; \vec{\mu}_k, \Sigma_k)}{\sum\limits_{j=1}^M w_jN(\vec{x}_{t} ; \vec{\mu}_k, \Sigma_k)} \label{eq:postr}\\
P(k) = & \frac{1}{T}\sum\limits_{i=1}^T Pr(k | \vec{x}_{t}) \label{eq:addms}
\end{align}
These soft-assignments are added over all $t$ to obtain the total mass of MFCC vectors belonging to the $k^{th}$ Gaussian (Eq \ref{eq:addms}). Normalization by $T$ is used to remove the effect of the duration of recordings. 
The final feature representation is $\vec{\alpha}^M=[P(1),..P(k)..P(M)]^T$. $\vec{\alpha}^M$ is an $M$-dimensional feature representation for a given recording. It captures how the MFCC vectors of a recording are distributed across the Guassians in $\mathcal{G}$. $\vec{\alpha}^M$ is normalized to sum to $1$ before using it for classifier training. 
\subsubsection{$\vec{\beta}$ features}
The next feature ($\vec{\beta}$), also based on the GMM $\mathcal{G}$, tries to capture the actual distribution of the MFCC vectors of a recording. This is done by adapting the parameters of $\mathcal{G}$ to the MFCC vectors of the recording using maximum {\em a posteriori} (MAP) estimation \cite{gauvain1994} \cite{bimbot}. The parameters $\vec{\mu}_k$ and $\vec{\sigma}_k$ of $k^{th}$ Gaussian is adapted according to Eq \ref{eq:mnmap} and \ref{eq:vrmap}.
\begin{align}
\hat{\vec{\mu}}_k=&\frac{n_k}{n_k+r}E_{k}(\vec{x})+\frac{r}{n_k+r}\vec{\mu_k} \label{eq:mnmap}\\
\hat{\vec{\sigma}}_k=&\frac{n_k}{n_k+r}E_{k}(\vec{x}^2)+\frac{r}{n_k+r}(\vec{\sigma}_k^2+\vec{\mu}_k^2) - \hat{\vec{\mu}}_k^2 \label{eq:vrmap}
\end{align}
The terms $n_k$, $E_{k}(\vec{x})$ and $E_{k}(\vec{x}^2)$ are computed according to Eq \ref{eq:totpost}-\ref{eq:vrexpt}. 
\begin{align}
n_{k}=&\sum\limits_{t=1}^T Pr(k | \vec{x}_{t}) \label{eq:totpost}\\
E_{k}(\vec{x})=&\frac{1}{n_{k}}\sum\limits_{t=1}^T Pr(k | \vec{x_{t}})\vec{x}_{t} \label{eq:mnexpt}\\
E_{k}(\vec{x}^2)=&\frac{1}{n_{k}}\sum\limits_{t=1}^T Pr(k | \vec{x_{t}})\vec{x}_{t}^2 \label{eq:vrexpt}
\end{align}

Eq \ref{eq:mnexpt} and \ref{eq:vrexpt} are the mean and variance estimates for MFCCs of a recording with respect to the background GMM $\mathcal{G}$. The relevance factor $r$ controls the effect of the original parameters on the new estimates. Although $w_k$ can also be adapted, their adaptation do not follow from general MAP estimation and hence we do not use weight updates. 

We obtain $4$ different feature representation using the adapted means ($\hat{\vec{\mu}}_k$) and variances ($\hat{\vec{\sigma}}_k$). The first one denoted by $\vec{\beta}^M$ is an $M \times D$ dimensional feature obtained by concatenating the adapted means $\hat{\vec{\mu}}_k$ for all $k$, that is $\vec{\beta}^M=[\hat{\vec{\mu}}_1^T,...\hat{\vec{\mu}}_K^T]^T$. $\vec{\beta}^M$ is a well-known feature for \emph{speaker verification} \cite{bimbot} and is usually referred as a ``Supervector'' feature. Campbell {\em et} {\em al. }\cite{campbell2006} showed that modifying the adapted means as $\hat{\vec{\mu}}_k^s=\sqrt{w_k}\Sigma_k^{-\frac{1}{2}}\hat{\vec{\mu}}_k$, makes the supervector features more suitable for linear SVMs. Since, $\Sigma_k$ is a diagonal matrix, this modification is a simple scaling in each dimension of $\hat{\vec{\mu}}_k$. The concatenation of $\hat{\vec{\mu}}_k^s$ as before gives us another $M \times D$ dimensional features denoted by $\vec{\beta}^M_s$. In general speaker verification uses only adapted means $\hat{\vec{\mu}}_k$. Here, we try to analyze the utility of adapted variance updates as well for audio event detection. We concatenate $\hat{\vec{\sigma}}_k$'s along with $\hat{\vec{\mu}}_k$ to obtain the third form of $\vec{\beta}$ features. This form denoted by $\vec{\beta}^M_{\sigma}$ is a $2 \times M \times D$ dimensional feature. The fourth and last form is obtained by concatenating scaled adapted means $\hat{\vec{\mu}}_k^s$ and $\hat{\vec{\sigma}}_k$ for all $k$. This feature is denoted by $\vec{\beta}^M_{s \sigma}$ and is again $2 \times M \times D$ dimensional. 
\subsection{Kernels}
As stated before we use SVM \cite{hastie2005elements} classifiers. We analyze different kernel functions for training SVMs on the above features. The first one is  linear kernel (LK). The other two kernels can be described in a general form for feature vectors $\vec{f}$ and $\vec{f}'$ by $K(\vec{f},\vec{f}')=\exp^{-\gamma D(\vec{f},\vec{f}')}$. For $D(\vec{f},\vec{f}')=||\vec{f}-\vec{f}'||^2_2$ we obtain the radial basis function kernel (RK). For histogram features ($\vec{\alpha}$ features) exponential Chi-square distance kernels (CK) are known to work well for vision tasks \cite{zhang2007local}. The Chi-square distance \cite{li2013sign} is given by $D(\vec{f},\vec{f}') = \sum_{i=1}^p \frac{(f_i-f'_i)^2}{f_i+f'_i}$, where $p$ is dimensionality of feature vectors $\vec{f}$. We use this kernel only for histogram features ($\vec{\alpha}$). Linear and RBF kernel are used for all features. It is worth noting that $\vec{\beta}^M_s$ have been designed specifically for linear kernels. In our experiments we try $\vec{\beta}^M_s$ with RBF kernels as well. 

\begin{table}[t]
\centering
\caption{Mean Average Precision for different cases (ESC-50)}
\label{tab:escap}
\resizebox*{1.0\columnwidth}{!}{
\begin{tabular}{|c|c|c|c|c|c|c|c|c|c|c|c|}
\hline  
 &\multicolumn{3}{c|}{$\vec{\alpha}^M$} &  \multicolumn{2}{c|}{$\vec{\beta}^M$} & \multicolumn{2}{c|}{$\vec{\beta}^M_s$} & \multicolumn{2}{c|}{$\vec{\beta}^M_{\sigma}$} & \multicolumn{2}{c|}{$\vec{\beta}^M_{s \sigma}$}\\ 
\cline{2-12}
$M$&LK&RK&CK&LK&RK&LK&RK&LK&RK&LK&RK\\
\hline 
32&0.165&0.337&0.462&0.439&0.46&0.519&0.495&0.332&0.390&0.322&0.375\\
\hline
64&0.247&0.418&0.558&0.416&0.423&0.523&0.513&0.325&0.340&0.321&0.333\\
\hline
128&0.302&0.46&0.611&0.416&0.411&0.528&0.530&0.292&0.290&0.296&0.294\\
\hline
256&0.339&0.463&0.622&0.380&0.377&0.523&0.521&0.276&0.268&0.274&0.267\\
\hline
\end{tabular}
}
\vspace{-0.10in}
\end{table}
\begin{table}[t]
\centering
\caption{Mean AUC for different cases (ESC-50)}
\label{tab:escauc}
\resizebox*{1.0\columnwidth}{!}{
\begin{tabular}{|c|c|c|c|c|c|c|c|c|c|c|c|}
\hline  
 &\multicolumn{3}{c|}{$\vec{\alpha}^M$} &  \multicolumn{2}{c|}{$\vec{\beta}^M$} & \multicolumn{2}{c|}{$\vec{\beta}^M_s$} & \multicolumn{2}{c|}{$\vec{\beta}^M_{\sigma}$} & \multicolumn{2}{c|}{$\vec{\beta}^M_{s \sigma}$}\\ 
\cline{2-12}
$M$&LK&RK&CK&LK&RK&LK&RK&LK&RK&LK&RK\\
\hline 
32&0.208&0.109&0.085&0.099&0.096&0.096&0.100&0.104&0.093&0.105&0.097\\
\hline
64&0.161&0.099&0.071&0.120&0.122&0.105&0.106&0.128&0.121&0.129&0.126\\
\hline
128&0.139&0.102&0.065&0.13&0.136&0.103&0.107&0.147&0.154&0.145&0.152\\
\hline
256&0.139&0.110&0.068&0.141&0.139&0.113&0.111&0.171&0.178&0.172&0.176\\
\hline
\end{tabular}
}
\vspace{-0.10in}
\end{table}
\vspace{-0.1in}
\section{Experiments and Results}
\vspace{-0.05in}
We report results on both \emph{UrbanSounds8k} and \emph{ESC-50} datasets. First, we divide the datasets into $10$ folds. UrbanSounds8k already comes with presorted folds and we follow the same fold structure. $9$ folds are used as training data and the $10$th fold is used for testing. This done in all $10$ ways such that each fold becomes test data once. This allows us to obtain results on the entire dataset and aggregate results over entire dataset are reported here. $20$-dimensional MFCC vectors are extracted for each audio recording using $30$ ms windows and a $15$ ms hop size. We experiment with $4$ different values of number of mixture components ($M$) in GMM $\mathcal{G}$. These are $32$, $64$, $128$, $256$. The relevance factor $r$ is set to $20$ for all experiments. For training SVMs we use LIBSVM \cite{chang2011libsvm}. The slack parameter in SVM training and the parameter $\gamma$ for RBF and Chi-Square kernel is obtained by performing cross validation over the given training data. 

We use two well known metrics for evaluation. The first one is Average Precision (AP) in the detection of each sound event, which is very relevant to retrieval scenarios. It is widely used in multimedia event detection \cite{2015trecvidover} challenges. The second one is miss probability vs false alarm rate curves. It is a more general evaluation method used in detection problems which is obtained by plotting miss probability and false alarm rate for different decision thresholds. The corresponding non-linear plot is referred as a DET curve \cite{martin1997det}. For DET curves, we use area under the curve (AUC) as a metric to characterize the curves. Since DET curves are error curves , the lower the AUC the better it is. Most results are reported in terms of mean average precision (MAP) and mean area under curves (MAUC). These are mean values of AP and AUC over all events of the corresponding dataset. More detailed results can be found on this webpage \cite{aedres}. For a few select cases we show AP and AUC values for all audio events as well. We start by showing results on the ESC-50 dataset. \\ \\
\textbf{ESC-50}: Table \ref{tab:escap} shows MAP values over all 50 events of ESC-50 . It shows MAP for different $M$ values, features and kernels. Table \ref{tab:escauc} shows the mean AUC values for different cases. We can observe that for $\vec{\alpha}$ features, exponential Chi-square kernel (CK) is much superior compared to linear (LK) and RBF (RK) kernels. There is an absolute improvement of $20$-$30\%$ in MAP using the exponential Chi-square kernel. The MAP also improves with increasing $M$. Among the $\vec{\beta}$ features $\vec{\beta}^M_s$ seems to be the best variant. Adding adapted variance $\hat{\vec{\sigma}}_k$ leads to inferior performance. Both $\vec{\beta}^M_{\sigma}$ and $\vec{\beta}^M_{s \sigma}$ perform poorly compared to other features. Between $\vec{\beta}^M_s$ and $\vec{\beta}^M$, the latter is clearly superior for any given kernel and $M$. An important point worth noting is that for $\vec{\beta}^M$ features the performance goes down as $M$ increases. This is not observed for $\vec{\beta}^M_s$ where MAP is more or less constant across all $M$. For $\vec{\beta}^M_s$ linear kernels are in general better than RBF kernels, although, the difference is small. 

If we look at the performance across all cases $\vec{\alpha}^M$ features with exponential Chi-square kernel are seen to result in the best performance, except for $M=32$. For $M=32$, MAP for $\vec{\beta}^M_s$ with LK is better by about $5\%$ compared to $\vec{\alpha}^M$ with CK. However, at higher $M$, $\vec{\alpha}^M$ is consistently better than $\vec{\beta}^M_s$ and at $M=256$ the gain is close to $10\%$ in absolute terms. Thus, if the number of Gaussians in $\mathcal{G}$ is not large it is expected that $\vec{\beta}^M_s$ features will lead to better results. The analysis of mean AUC measure is more or less similar to MAP. A few points are worth noting. In terms of area under error curves $\vec{\alpha}^M$ with CK is better than others across all $M$. Mean AUC with $\vec{\beta}^M_s$ is again more or less constant for different $M$ and kernels.
\begin{figure}[t] 
\centering
\includegraphics[width=0.48\columnwidth,height=1.0in]{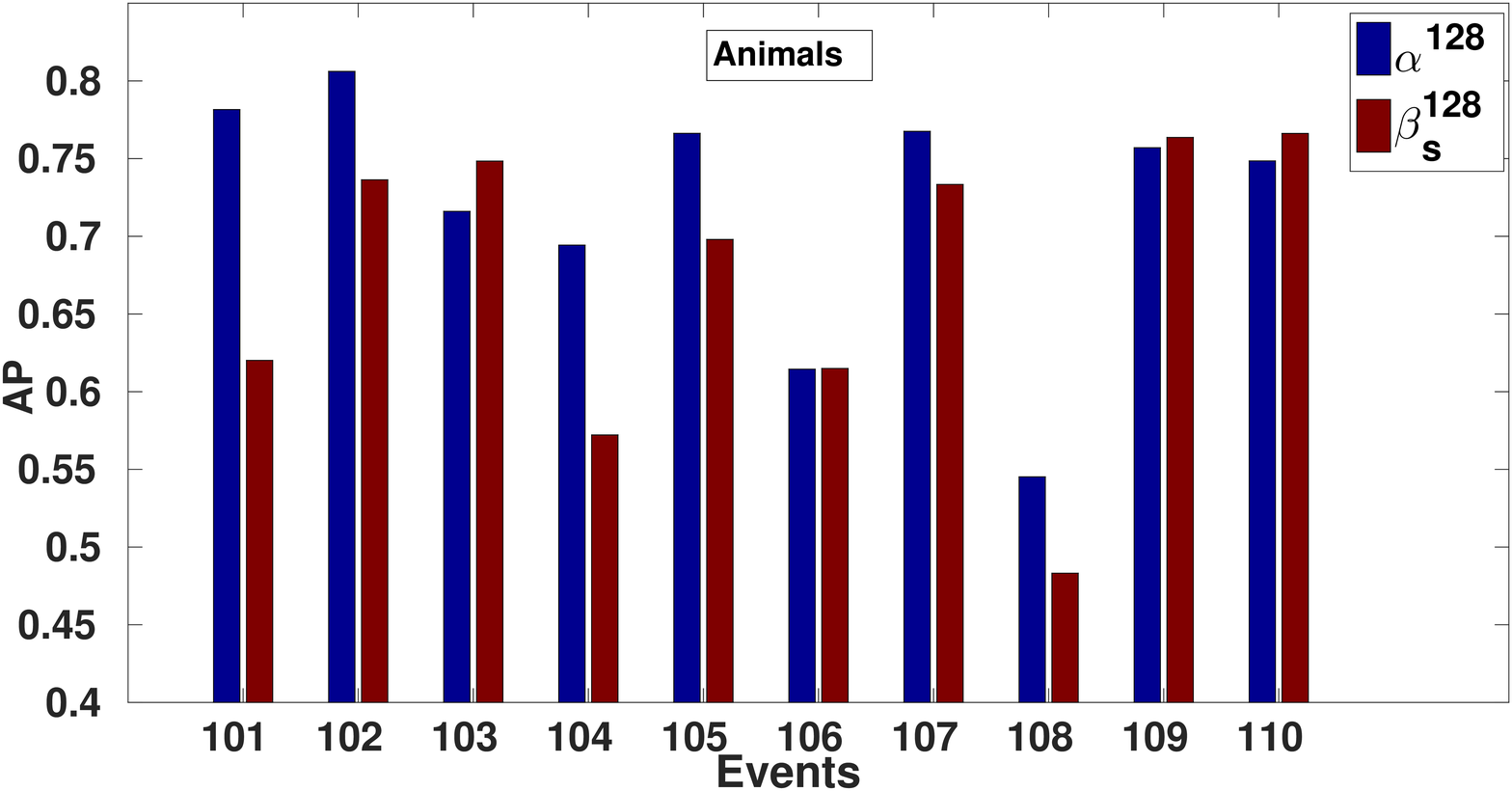}
\includegraphics[trim=0.5in 0.0in 0.5in 0.5in,width=0.48\columnwidth,height=1.0in]{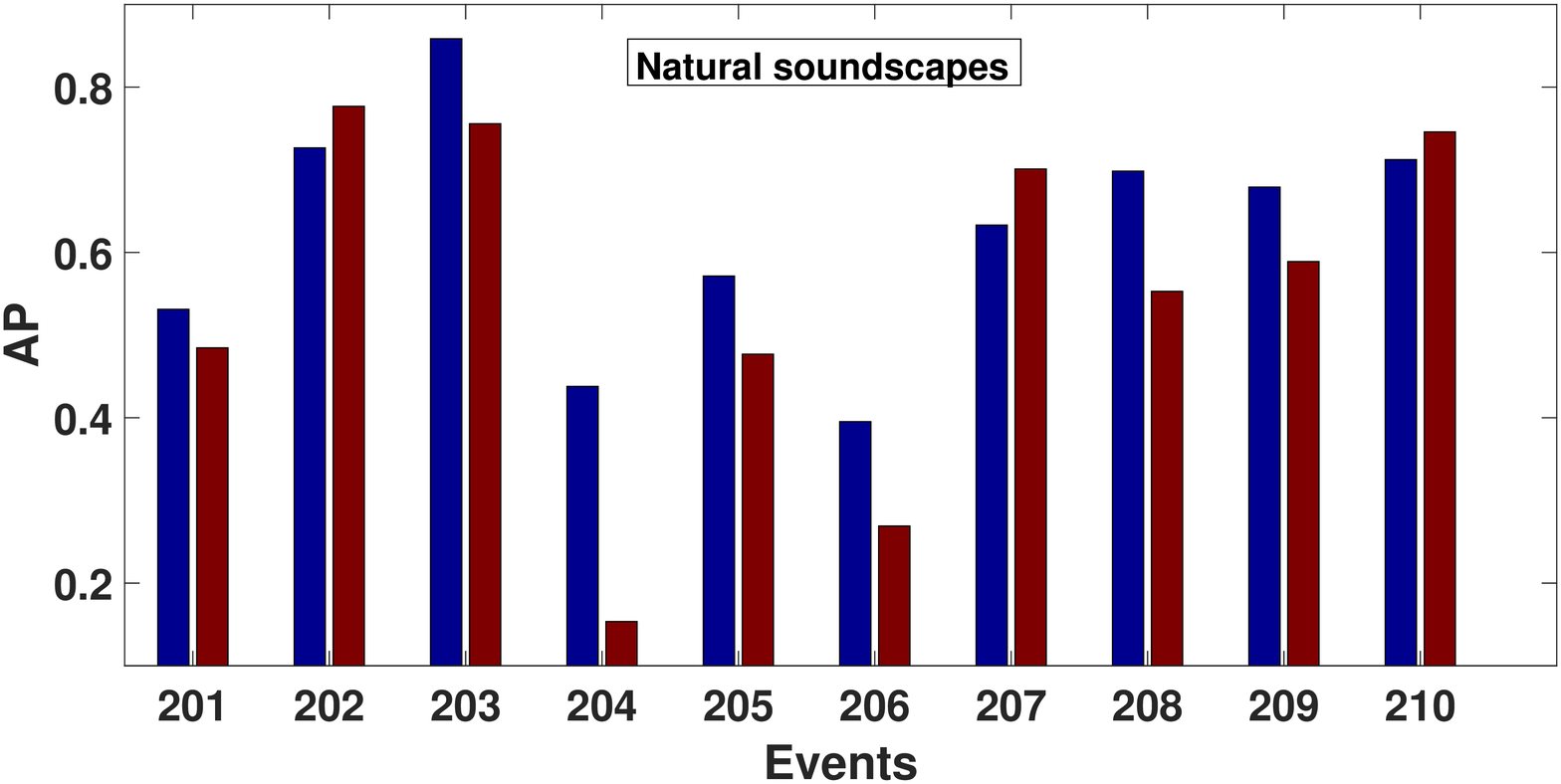}
\includegraphics[width=0.48\columnwidth,height=1.0in]{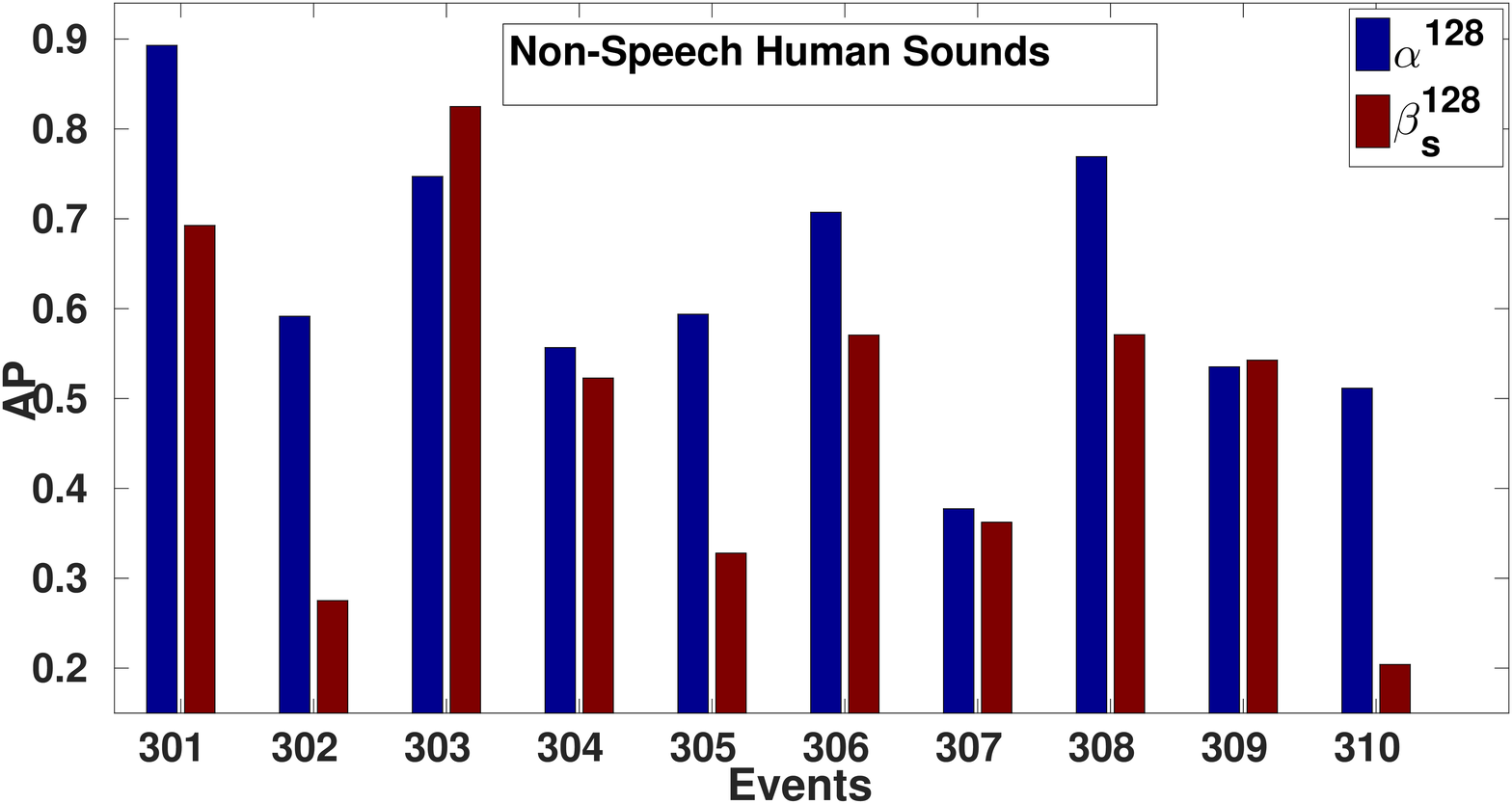}
\includegraphics[trim=0.5in 0.0in 0.5in 0.5in,width=0.48\columnwidth,height=1.0in]{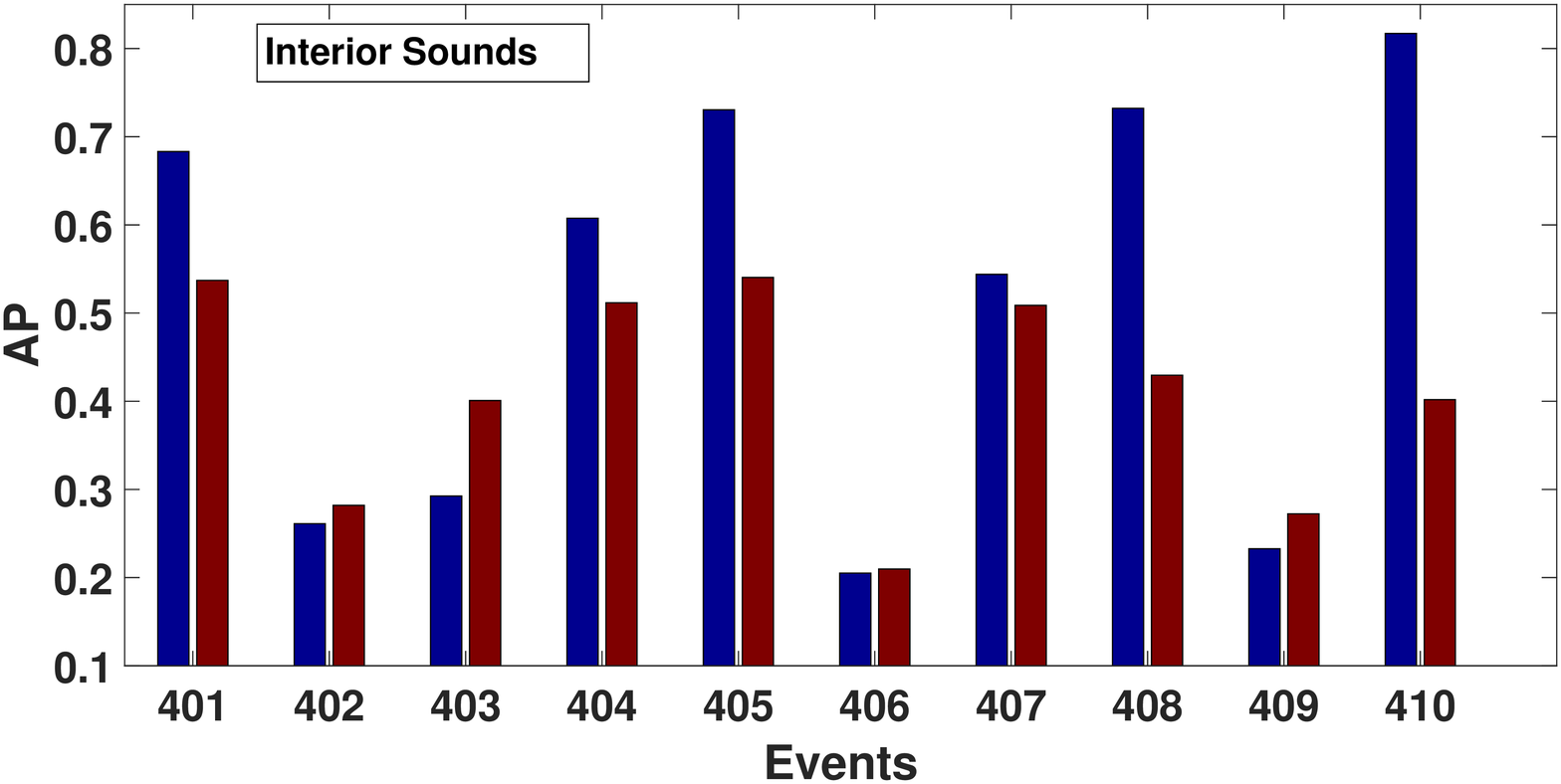}
\includegraphics[width=0.48\columnwidth,height=1.0in]{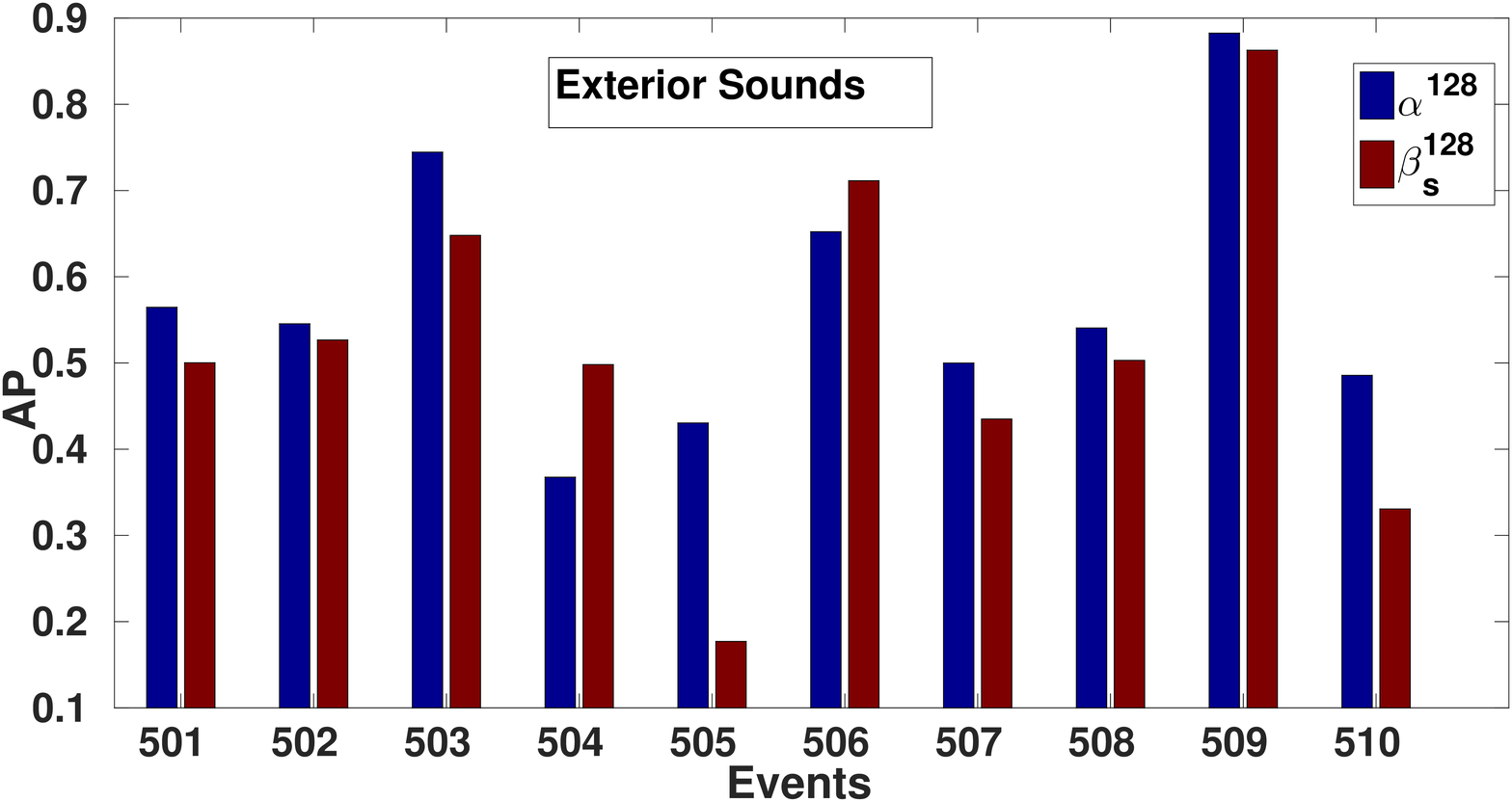}
\includegraphics[width=0.48\columnwidth,height=1.0in]{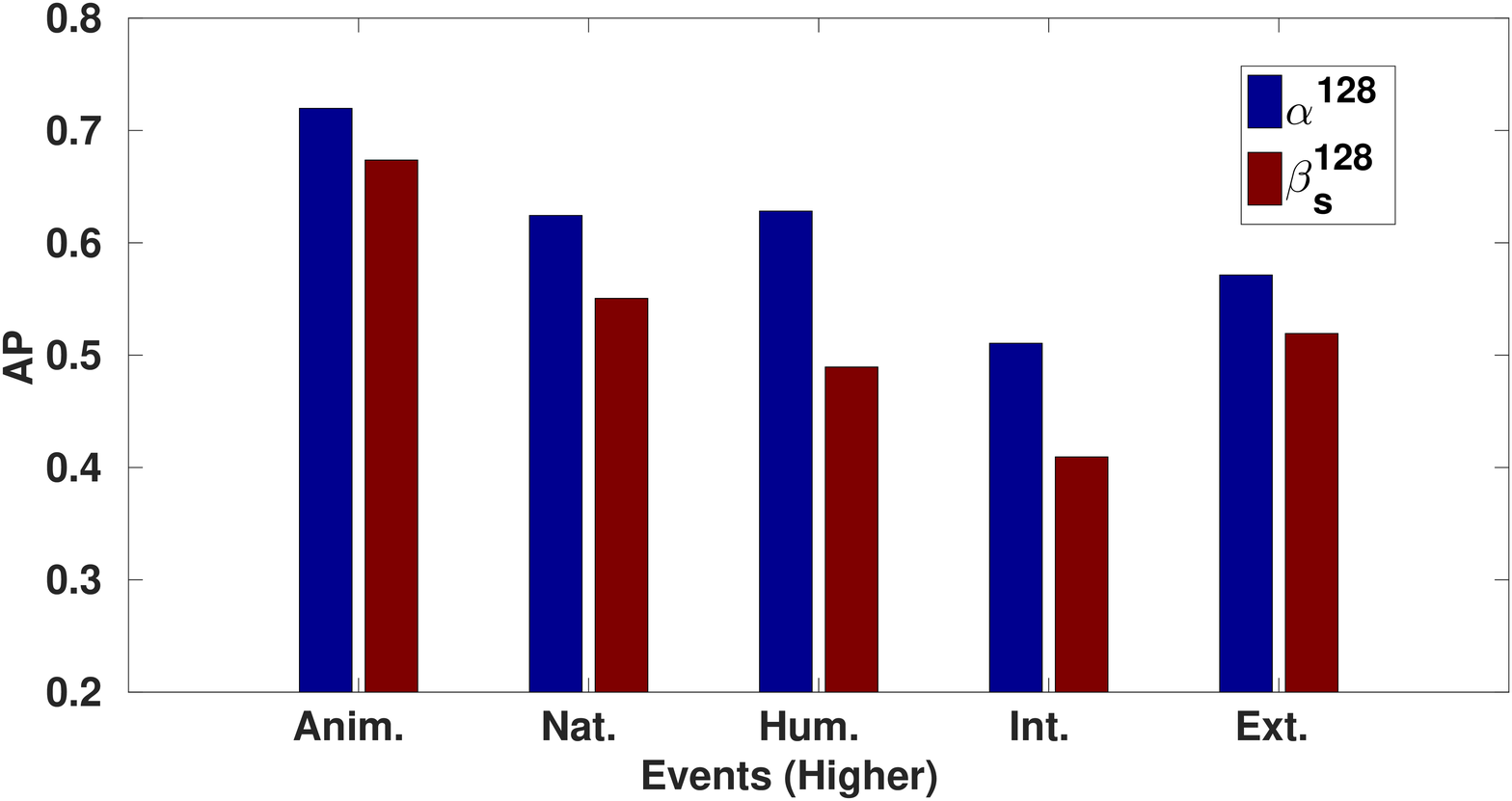}
\caption{ESC-50:Event AP values for $\vec{\alpha}^{128}$ + CK (BLUE)  and $\vec{\beta}^{128}_s$ + LK (RED), Bottom Right Corner - MAP over Higher Semantic Categories}
\label{fig:escevt}
\end{figure}

We show event-wise AP values for all $50$ events in Figure \ref{fig:escevt}. The events have been grouped according their higher semantic categories which are animals (Anim), natural soundscapes(Nat), Non-speech human sounds (Hum), Interior Sounds (Int) and Exterior Sounds (Ext). For convenience, the acoustic events of each higher semantic category have been referred by their indices such as $101$ to $110$ for animals and so on. The exact name of the event can be found here \cite{aedres}. Among events in the \emph{Animal} category \emph{insect sounds} (108) are most difficult to detect. Among the \emph{Interior} sound categories events such as Washing Machine (406) and Clock Ticking (409) are particularly difficult. Clock tick event is surprisingly difficult even though due it has some unique characteristics from the perspective of human listener and hence we would expect to do better on it. Fireworks(509), Siren(503) and Trains (506) in the Exterior sound sets can be easily detected. Also, there are events such as Wind(207), Thunderstorm(210), Clapping(303), Train(506)  where  $\vec{\beta}^M_s$  features do better than $\vec{\alpha}^M$. 

Figure \ref{fig:escevt} (bottom right) also shows MAP over each higher semantic category. It can be observed that \emph{animal} sounds are the easiest to detect and \emph{interior sounds} are the hardest. The maximum difference between $\vec{\alpha}$ and $\vec{\beta}$ features is for non-speech human sounds. Mean AUC results has more or less similar trend. 

\begin{table}[t]
\centering
\caption{Mean AP for different cases (UrbanSounds8k)}
\resizebox*{1.0\columnwidth}{!}{
\begin{tabular}{|c|c|c|c|c|c|c|c|}
\hline  
 &\multicolumn{3}{c|}{$\vec{\alpha}^M$} &  \multicolumn{2}{c|}{$\vec{\beta}^M$} & \multicolumn{2}{c|}{$\vec{\beta}^M_s$}\\ 
\cline{2-8}
$M$&LK&RK&CK&LK&RK&LK&RK\\
\hline 
32&0.361&0.493&0.574&0.514&0.541&0.536&0.528\\
\hline
64&0.42&0.528&0.580&0.456&0.506&0.506&0.513\\
\hline
128&0.49&0.533&0.614&0.392&0.448&0.474&0.499\\
\hline
256&0.518&0.554&0.624&0.323&0.377&0.419&0.454\\
\hline
\end{tabular}
}
\label{tab:us8kap}
\vspace{-0.15in}
\end{table}
\begin{table}[t]
\centering
\caption{Mean AUC for different cases (UrbanSounds8k)}
\resizebox*{1.0\columnwidth}{!}{
\begin{tabular}{|c|c|c|c|c|c|c|c|}
\hline  
 &\multicolumn{3}{c|}{$\vec{\alpha}^M$} &  \multicolumn{2}{c|}{$\vec{\beta}^M$} & \multicolumn{2}{c|}{$\vec{\beta}^M_s$}\\ 
\cline{2-8}
$M$&LK&RK&CK&LK&RK&LK&RK\\
\hline 
32&0.241&0.176&0.148&0.177&0.165&0.168&0.174\\
\hline
64&0.199&0.166&0.140&0.219&0.195&0.187&0.188\\
\hline
128&0.183&0.167&0.138&0.270&0.229&0.206&0.200\\
\hline
256&0.193&0.175&0.142&0.320&0.270&0.238&0.206\\
\hline
\end{tabular}
}
\label{tab:us8kauc}
\vspace{-0.15in}
\end{table}
\textbf{UrbanSounds8k}: Table \ref{tab:us8kap} and Table \ref{tab:us8kauc} shows MAP  and MAUC values respectively on the UrbanSounds8k dataset for different features and kernels.  $\vec{\beta}^M_{\sigma}$ and $\vec{\beta}^M_{s \sigma}$ have been removed from analysis due to their inferior performance. Once again we notice that $\vec{\alpha}$ features with the exponential Chi-square kernel is in general the best feature and kernel pair. Interestingly, in this case $\vec{\alpha}^M$ with CK is better than all $\vec{\beta}$ features for lower $M$ values as well. At low $M$ MAP gain is between $3-8\%$ whereas for large $M$ the difference goes upto $16-17\%$ in absolute terms. The primary reason for this is the fact that for $\vec{\beta}$ features performance goes down as we increase $M$ whereas for $\vec{\alpha}$ features the performance goes up. Among the $\vec{\beta}$ features, $\vec{\beta}^M_s$ is in general better than $\vec{\beta}^M$, though, for $M=32$ and $M=64$ under suitable kernel both features give more or less similar performance. However, we had previously observed for the ESC-50 dataset that $\vec{\beta}^M_s$ is always better than $\vec{\beta}^M$. Another new observation for $\vec{\beta}^M_s$ on the UrbanSounds8k dataset is that MAP goes down (MAUC goes up) with increasing $M$ which was not the case in ESC-50. 

Mean AUC values also offers some interesting observations. For $\vec{\alpha}$ features with CK, the MAUC values more or less remain constant. So, although increasing $M$ helps in obtaining better ranked outputs (MAP increases by about $5\%$), the detection across the entire dataset (AUC) generally stays constant. Thus higher $M$ is especially helpful for $\vec{\alpha}$ features if we are in a retrieval scenario where the recordings belonging to a certain class are to be retrieved from a larger set. 

\begin{figure}[t] 
\centering
\includegraphics[width=0.48\columnwidth,height=1.0in]{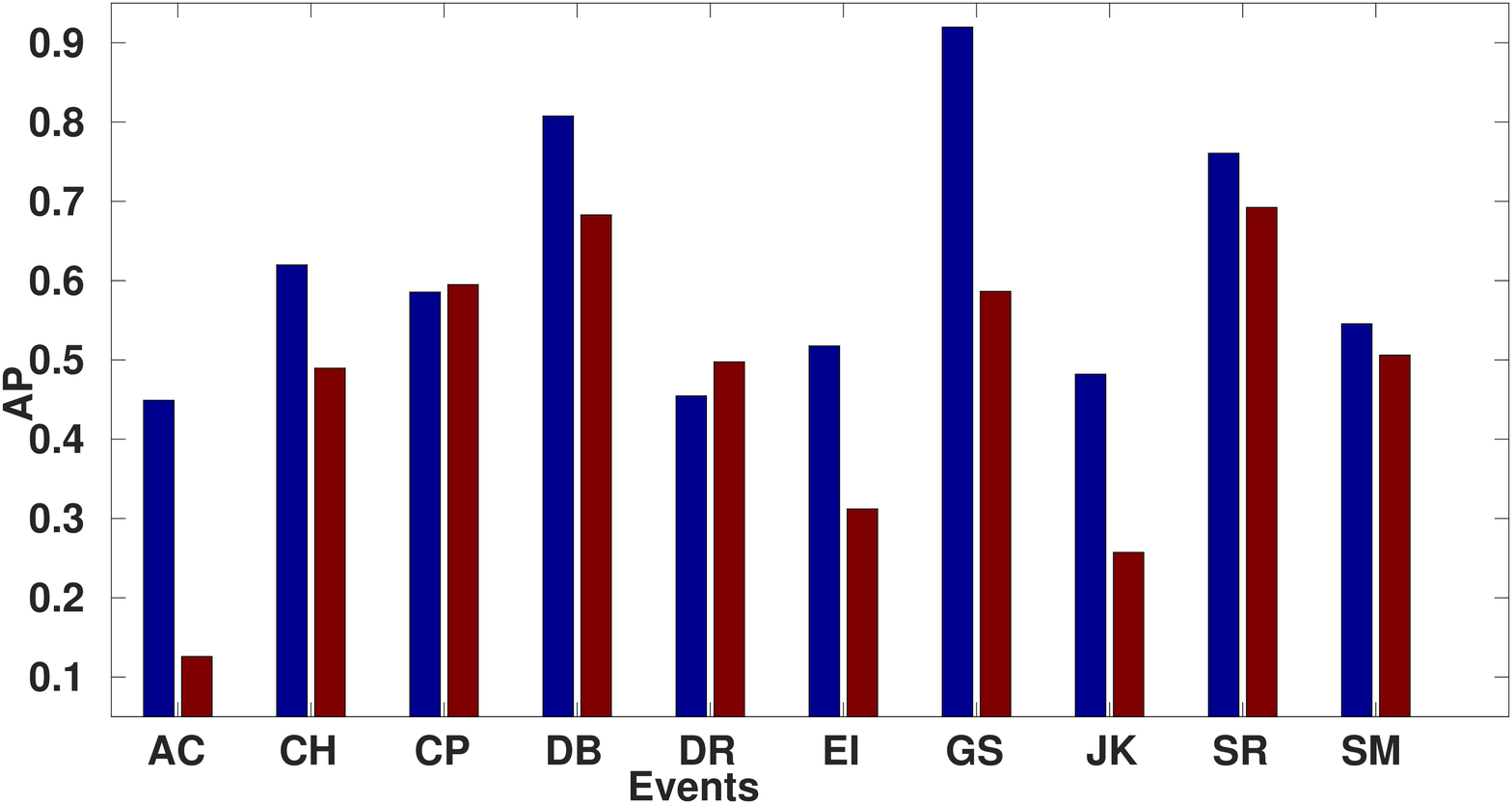}
\includegraphics[width=0.48\columnwidth,height=1.0in]{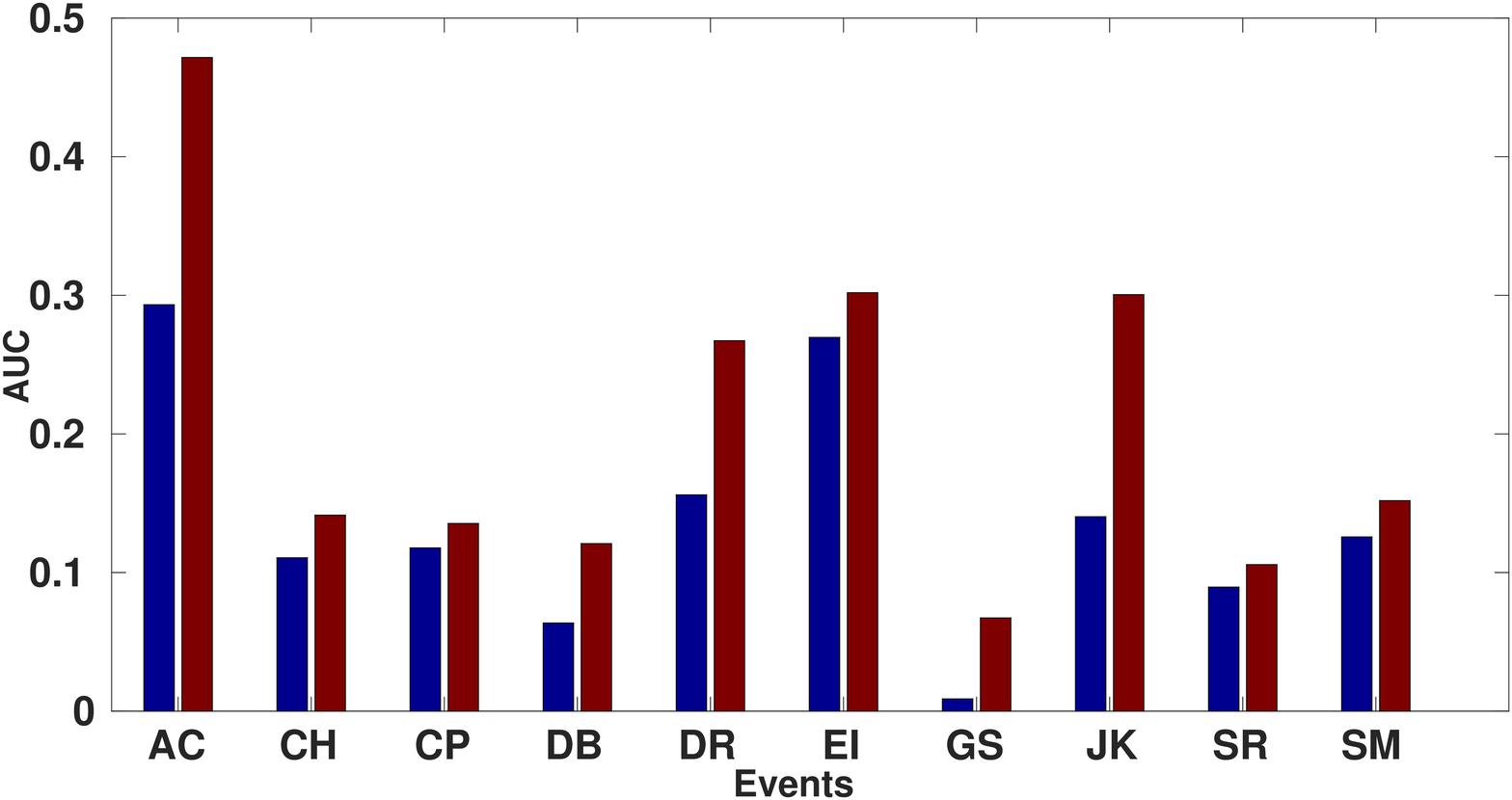}
\caption{UrabnSounds8k: Left - Event AP values, Right - Event AUC values ($\vec{\alpha}^{128}$ + CK (BLUE)  and $\vec{\beta}^{128}_s$ + LK (RED)) }
\label{fig:us8kevt}
\vspace{-0.05in}
\end{figure}

Event wise results (AP and AUC) for $10$ events of UrbanSounds8k is shown in Figure \ref{fig:us8kevt}. The event names have been encoded in the figure as per the description in Section \ref{sec:dtst}. Events like \emph{dog barking}, \emph{Siren} and \emph{Gun Shot} are again easy do detect. 

\begin{figure}[t!] 
\centering
\includegraphics[width=0.48\columnwidth,height=1.0in]{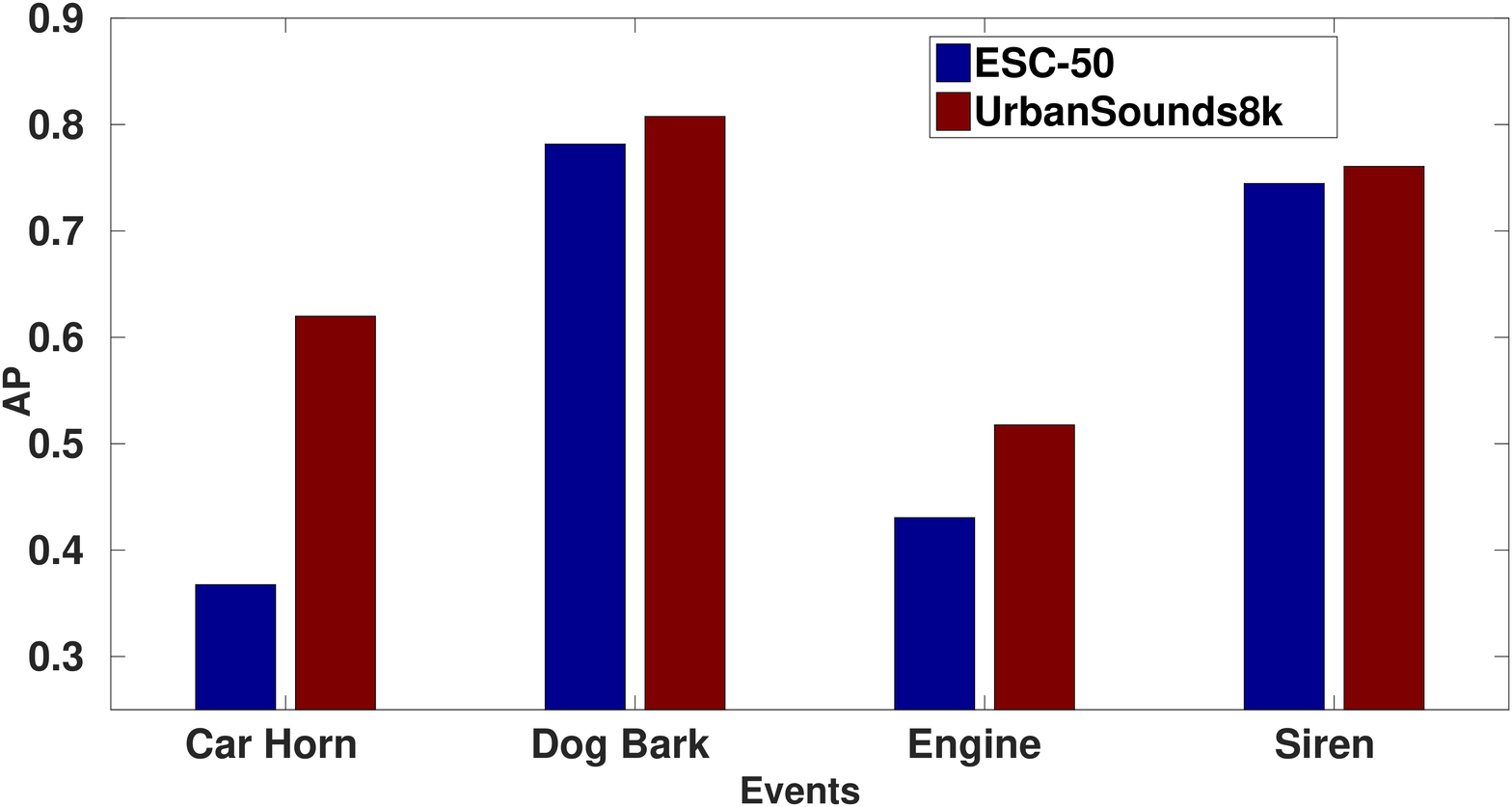}
\includegraphics[width=0.48\columnwidth,height=1.0in]{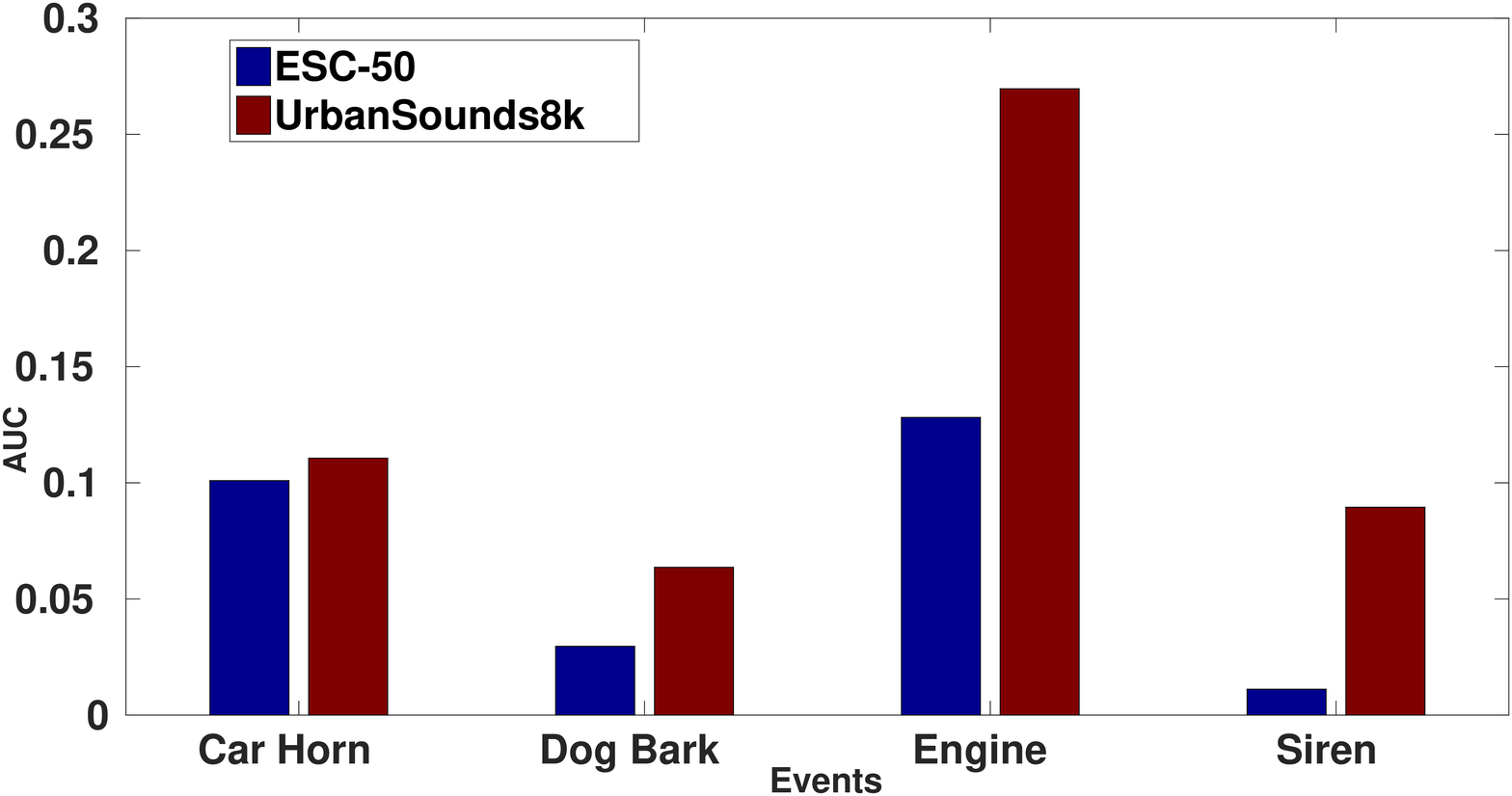}
\caption{ESC-50 and UrbanSounds8k: Left - Event AP values, Right - Event AUC values ($\vec{\alpha}^{128}$ + CK for both datasets) }
\label{fig:bothevt}
\end{figure}

The source of the original audios for both datasets is Freesound \cite{fsound}.  This allows us to make general comments for events which are common to both datasets, namely, \emph{Car Horn, Dog bark, Engine} and \emph{Siren}. The results are shown in Figure \ref{fig:bothevt}. We observe that we do well on Dog barking on both datasets whereas complex events such as Engine are inherently difficult to detect and poor performance is obtained on both datasets. 

\section{Conclusions}
\vspace{-0.05in}
In this work we have performed a comprehensive analysis of audio event detection using two publicly available datasets. Overall, we presented results for $56$ unique events across the two datasets. Our analysis included two broad sets of features and overall $5$ different feature types. The features are GMM-based representations. The classifier side included SVMs with $3$ different kernel types. Our overall analysis on both datasets favors use of exponential Chi-square kernels over $\vec{\beta}$ features. However, given the variation we observed over different cases it is possible that the fusion of these methods might actually lead to much superior results. The fusion can be done at either feature level where $\vec{\alpha}^M$ and $\vec{\beta}^M_s$ features are concatenated for each recording and classified with any appropriate classifier, or at a decision level that combines individual classifier decisions. Our analysis shows that $\vec{\alpha}^M$ features work remarkably well with exponential Chi-square kernels, suggesting that fusion at decision level might be the better choice. In this case the score(probability) outputs for a test recording from each system can be combined by simple averaging to obtain the final score(probability) for that recording. The relevance factor $r$ in computating $\vec{\beta}$ features controls the contribution of means and variances of GMM $\mathcal{G}$ to the features. Its possible that fine tuning $r$ may improve results for $\vec{\beta}$ features. 

ESC-50 dataset also allowed us to study audio event detection at lower and higher semantic levels. We observed that events in the Animal group are in general easier to detect. Interior sound events which includes events such as Washing Machine, Vacuum Cleaner, Clock Ticking are relatively harder to detect. Together both datasets allowed us to study audio event detection in an exhaustive manner for the features and kernels used in this work. This is potentially very helpful in standardizing AED methods and results. 
\newpage
\eightpt
\bibliographystyle{IEEEtran}
\bibliography{refs}

\begin{thebibliography}{10}
\providecommand{\url}[1]{#1}
\csname url@samestyle\endcsname
\providecommand{\newblock}{\relax}
\providecommand{\bibinfo}[2]{#2}
\providecommand{\BIBentrySTDinterwordspacing}{\spaceskip=0pt\relax}
\providecommand{\BIBentryALTinterwordstretchfactor}{4}
\providecommand{\BIBentryALTinterwordspacing}{\spaceskip=\fontdimen2\font plus
\BIBentryALTinterwordstretchfactor\fontdimen3\font minus
  \fontdimen4\font\relax}
\providecommand{\BIBforeignlanguage}[2]{{%
\expandafter\ifx\csname l@#1\endcsname\relax
\typeout{** WARNING: IEEEtran.bst: No hyphenation pattern has been}%
\typeout{** loaded for the language `#1'. Using the pattern for}%
\typeout{** the default language instead.}%
\else
\language=\csname l@#1\endcsname
\fi
#2}}
\providecommand{\BIBdecl}{\relax}
\BIBdecl

\bibitem{2015trecvidover}
P.~Over, G.~Awad, M.~Michel, J.~Fiscus, W.~Kraaij, A.~Smeaton, G.~Quéenot, and
  R.~Ordelman, ``Trecvid 2015 -- an overview of the goals, tasks, data,
  evaluation mechanisms and metrics,'' in \emph{Proceedings of TRECVID
  2015}.\hskip 1em plus 0.5em minus 0.4em\relax NIST, USA, 2015.

\bibitem{1}
G.~Valenzise, L.~Gerosa, M.~Tagliasacchi, F.~Antonacci, and A.~Sarti, ``Scream
  and gunshot detection and localization for audio-surveillance systems,'' in
  \emph{Advanced Video and Signal Based Surveillance, {IEEE} Conference on},
  2007, pp. 21--26.

\bibitem{stowell2014}
D.~Stowell and M.~D. Plumbley, ``Automatic large-scale classification of bird
  sounds is strongly improved by unsupervised feature learning,'' \emph{PeerJ},
  2014.

\bibitem{ruizmultiple}
J.~Ruiz-Mu{\~n}oz, M.~Orozco-Alzate, and G.~Castellanos-Dominguez, ``Multiple
  instance learning-based birdsong classification using unsupervised recording
  segmentation,'' in \emph{Proceedings of the 24th Intl. Conf. on Artificial
  Intelligence}, 2015.

\bibitem{battaglino2015}
D.~Battaglino, A.~Mesaros, L.~Lepauloux, L.~Pilati, and N.~Evans, ``Acoustic
  context recognition for mobile devices using a reduced complexity svm,'' in
  \emph{Signal Processing Conference (EUSIPCO), 2015 23rd European}.\hskip 1em
  plus 0.5em minus 0.4em\relax IEEE, 2015, pp. 534--538.

\bibitem{eronen}
A.~J. Eronen, V.~T. Peltonen, J.~T. Tuomi, A.~P. Klapuri, S.~Fagerlund,
  T.~Sorsa, G.~Lorho, and J.~Huopaniemi, ``Audio-based context recognition,''
  \emph{Audio, Speech, and Language Processing, IEEE Transactions on}, vol.~14,
  no.~1, pp. 321--329, 2006.

\bibitem{zhuang2010}
X.~Zhuang, X.~Zhou, M.~A. Hasegawa-Johnson, and T.~S. Huang, ``Real-world
  acoustic event detection,'' \emph{Pattern Recognition Letters}, vol.~31,
  no.~12, pp. 1543--1551, 2010.

\bibitem{7}
S.~Pancoast and M.~Akbacak, ``Bag-of-audio-words approach for multimedia event
  classification,'' in \emph{Interspeech}, 2012.

\bibitem{supbow}
A.~Plinge, R.~Grzeszick, G.~Fink \emph{et~al.}, ``A bag-of-features approach to
  acoustic event detection,'' in \emph{IEEE ICASSP}, 2014.

\bibitem{kumar2013event}
A.~Kumar, R.~Hegde, R.~Singh, and B.~Raj, ``Event detection in short duration
  audio using gaussian mixture model and random forest classifier,'' in
  \emph{Signal Processing Conference (EUSIPCO), 2013 Proceedings of the 21st
  European}.\hskip 1em plus 0.5em minus 0.4em\relax IEEE, 2013, pp. 1--5.

\bibitem{foggia2015reliable}
P.~Foggia, N.~Petkov, A.~Saggese, N.~Strisciuglio, and M.~Vento, ``Reliable
  detection of audio events in highly noisy environments,'' \emph{Pattern
  Recognition Letters}, vol.~65, pp. 22--28, 2015.

\bibitem{Ashraf2015}
K.~Ashraf, B.~Elizalde, F.~Iandola, M.~Moskewicz, J.~Bernd, G.~Friedland, and
  K.~Keutzer, ``Audio-based multimedia event detection with {DNNs} and sparse
  sampling,'' in \emph{Proc. of the 5th ACM International Conference on
  Multimedia Retrieval}, 2015.

\bibitem{gencoglu}
O.~Gencoglu, T.~Virtanen, and H.~Huttunen, ``Recognition of acoustic events
  using deep neural networks,'' in \emph{Signal Processing Conference
  (EUSIPCO), 2014 Proceedings of the 22nd European}.\hskip 1em plus 0.5em minus
  0.4em\relax IEEE, 2014, pp. 506--510.

\bibitem{12}
A.~Kumar, P.~Dighe, R.~Singh, S.~Chaudhuri, and B.~Raj, ``Audio event detection
  from acoustic unit occurrence patterns,'' in \emph{IEEE ICASSP}, 2012, pp.
  489--492.

\bibitem{specex}
X.~Lu, Y.~Tsao, S.~Matsuda, and C.~Hori, ``Sparse representation based on a bag
  of spectral exemplars for acoustic event detection,'' in \emph{IEEE ICASSP},
  2014, pp. 6255--6259.

\bibitem{mesaros2015sound}
A.~Mesaros, T.~Heittola, O.~Dikmen, and T.~Virtanen, ``Sound event detection in
  real life recordings using coupled matrix factorization of spectral
  representations and class activity annotations,'' in \emph{ICASSP}.\hskip 1em
  plus 0.5em minus 0.4em\relax IEEE, 2015, pp. 606--618.

\bibitem{kumar2016audio}
A.~Kumar and B.~Raj, ``Audio event detection using weakly labeled data,'' in
  \emph{24th ACM International Conference on Multimedia}.\hskip 1em plus 0.5em
  minus 0.4em\relax ACM Multimedia, 2016.

\bibitem{anuragweakly}
------, ``Weakly supervised scalable audio content analysis,'' in \emph{2016
  IEEE International Conference on Multimedia and Expo (ICME)}.\hskip 1em plus
  0.5em minus 0.4em\relax IEEE, 2016.

\bibitem{pascal-voc-2011}
M.~Everingham, L.~Van~Gool, C.~K.~I. Williams, J.~Winn, and A.~Zisserman, ``The
  {PASCAL} {V}isual {O}bject {C}lasses {C}hallenge 2011 {(VOC2011)}
  {R}esults,''
  http://www.pascal-network.org/challenges/VOC/voc2011/workshop/index.html.

\bibitem{krizhevsky2009learning}
A.~Krizhevsky and G.~Hinton, ``Learning multiple layers of features from tiny
  images,'' 2009.

\bibitem{deng2009imagenet}
J.~Deng, W.~Dong, R.~Socher, L.-J. Li, K.~Li, and L.~Fei-Fei, ``Imagenet: A
  large-scale hierarchical image database,'' in \emph{Computer Vision and
  Pattern Recognition, 2009. CVPR 2009. IEEE Conference on}.\hskip 1em plus
  0.5em minus 0.4em\relax IEEE, 2009, pp. 248--255.

\bibitem{salamon2014dataset}
J.~Salamon, C.~Jacoby, and J.~P. Bello, ``A dataset and taxonomy for urban
  sound research,'' in \emph{Proceedings of the ACM International Conference on
  Multimedia}.\hskip 1em plus 0.5em minus 0.4em\relax ACM, 2014, pp.
  1041--1044.

\bibitem{piczak2015esc}
K.~J. Piczak, ``Esc: Dataset for environmental sound classification,'' in
  \emph{Proceedings of the 23rd Annual ACM Conference on Multimedia
  Conference}.\hskip 1em plus 0.5em minus 0.4em\relax ACM, 2015, pp.
  1015--1018.

\bibitem{mesaros2016tut}
A.~Mesaros, T.~Heittola, and T.~Virtanen, ``Tut database for acoustic scene
  classification and sound event detection,'' in \emph{24th European Signal
  Processing Conference}, 2016.

\bibitem{aedres}
A.~Kumar. Aed. \url{http://www.cs.cmu.edu/%7Ealnu/AED.htm} Copy and Paste in
  browser if clicking does not work.

\bibitem{gauvain1994}
J.~Gauvain and C.~Lee, ``Maximum a posteriori estimation for multivariate
  gaussian mixture observations of markov chains,'' \emph{Speech and audio
  processing, IEEE Trans. on}, 1994.

\bibitem{bimbot}
F.~Bimbot and et~al., ``A tutorial on text-independent speaker verification,''
  \emph{EURASIP journal on applied signal processing}, vol. 2004, pp. 430--451,
  2004.

\bibitem{campbell2006}
W.~Campbell, D.~Sturim, and A.~Reynolds, ``Support vector machines using gmm
  supervectors for speaker verification,'' \emph{Signal Processing Letters,
  IEEE}, pp. 308--311, 2006.

\bibitem{hastie2005elements}
T.~Hastie, R.~Tibshirani, J.~Friedman, and J.~Franklin, ``The elements of
  statistical learning: data mining, inference and prediction,'' \emph{The
  Mathematical Intelligencer}, vol.~27, no.~2, pp. 83--85, 2005.

\bibitem{zhang2007local}
J.~Zhang, M.~Marsza{\l}ek, S.~Lazebnik, and C.~Schmid, ``Local features and
  kernels for classification of texture and object categories: A comprehensive
  study,'' \emph{International journal of computer vision}, vol.~73, no.~2, pp.
  213--238, 2007.

\bibitem{li2013sign}
P.~Li, G.~Samorodnitsk, and J.~Hopcroft, ``Sign cauchy projections and
  chi-square kernel,'' in \emph{Advances in Neural Information Processing
  Systems}, 2013, pp. 2571--2579.

\bibitem{chang2011libsvm}
C.-C. Chang and C.-J. Lin, ``Libsvm: a library for support vector machines,''
  \emph{ACM Transactions on Intelligent Systems and Technology (TIST)}, vol.~2,
  no.~3, p.~27, 2011.

\bibitem{martin1997det}
A.~Martin, G.~Doddington, T.~Kamm, M.~Ordowski, and M.~Przybocki, ``The det
  curve in assessment of detection task performance,'' DTIC Document, Tech.
  Rep., 1997.

\bibitem{fsound}
FreeSound, ``\url{https://freesound.org/},'' 2015.

\end{thebibliography}


\end{document}